\documentclass[prb,amsmath,twocolumn,superscriptaddress,showpacs]{revtex4}

\usepackage{amsfonts}
\usepackage{amsmath}
\usepackage{amssymb}
\usepackage{graphicx}
\usepackage{txfonts}

\renewcommand{\vec}[1]{\mathbf{#1}}
\newcommand{\tx}{{\tilde{x}}}

\begin{document}
	\title{Missing Shapiro steps and the $8\pi$-periodic Josephson effect\\
	in interacting helical electron systems}
	\author{Christopher J. Pedder}
	\affiliation{Physics and Materials Science Research Unit, University of Luxembourg, L-1511 Luxembourg}
	\author{Tobias Meng}
	\affiliation{Institut f\"ur Theoretische Physik, Technische Universit\"at Dresden, 01062 Dresden, Germany}
	\author{Rakesh P. Tiwari}
	\affiliation{Department of Physics, University of Basel, Klingelbergstrasse 82, CH-4056 Basel, Switzerland}
	\affiliation{Department of Physics, McGill University, Montr\'eal, Qu\'ebec, Canada H3A 2T8}
	\author{Thomas L. Schmidt}
	\email{thomas.schmidt@uni.lu}
	\affiliation{Physics and Materials Science Research Unit, University of Luxembourg, L-1511 Luxembourg}
	
	\date{\today}
	
	\begin{abstract}
	Two-particle backscattering in time-reversal invariant interacting helical electron systems can lead to the formation of quasiparticles with charge $e/2$. We propose a way to detect such states by means of the Josephson effect in the presence of proximity-induced superconductivity. In this case, the existence of $e/2$ charges leads to an $8 \pi$-periodic component of the Josephson current which can be identified through measurement of Shapiro steps in Josephson junctions. In particular, we show that even when there is weak explicit time-reversal symmetry breaking, which causes the two-particle backscattering to be a sub-leading effect at low energies, its presence can still be detected in driven, current-biased Shapiro step measurements. The disappearance of some of these steps as a function of the drive frequency is directly related to the existence of non-Abelian zero-energy states. We suggest that this effect can be measured in current state-of-the-art Rashba wires.
	\end{abstract}
	
	
	\maketitle
	
\section{Introduction}
One of the most promising experimental avenues for the discovery of Majorana fermions \cite{Nayak+2008,leijnse12,Alicea+2012,beenakker13, majorana37} as quasiparticles in solid-state systems \cite{read00,Ivanov+2001,Kitaev+2001,Fu+2008,Fu+2009,Sato+2009,Sato+2009B,Oreg+2010,Lutchyn+2010} has been quantum wires with spin-orbit coupling. In such one-dimensional (1D) wires, the combined effects of Rashba spin-orbit coupling (RSOC) or helical nuclear order,\cite{Braunecker+09b,Braunecker+10,Braunecker+15} proximity-coupling to a superconductor, and an externally applied magnetic field \cite{Oreg+2010,Lutchyn+2010} open a band gap in the single-particle spectrum of the wire. This gap only closes at the ends of the wire, leading to localized zero-energy states which are believed to behave as Majorana fermions. Related, but more exotic bound states have been predicted to descend from strongly correlated phases which display charge fractionalization, such as the edges of fractional quantum Hall systems.\cite{fendley12,Lindner+2012,Clarke+2013,alicea15} or fractionalized helical states.\cite{Meng+14b,Oreg+2014}

However, unambiguous confirmation of such Majorana quasiparticles remains elusive. Several experimental systems have shown signatures that are consistent with MBS,\cite{Deng+2012,Mourik+2012,Wesperen+2013,Nadj+2014} but a common objection is that such tests are not directly sensitive to the specific properties of Majorana states. Recently, more targeted experimental tests have been used to rule out other possible causes. Measured current-voltage curves for proximitized edge states of HgTe, and nanowires with RSOC have shown the disappearance of odd numbered ``Shapiro steps''.\cite{rokhinson12,Maier+2015,Wiedenmann+2015} The disappearance of the odd steps is consistent with theoretical predictions\cite{Dominguez+2012,Dominguez+17} that there are tunnelling states with charge $e$, rather than the usual $2e$ for Cooper pairs, leading to a $4\pi$-periodic component of the Josephson current.

Crucial to the localization of bound states in these Majorana systems is the opening of a gap in the spectrum at the Dirac point at $k=0$, see Fig.~\ref{Interactions}. This so-called \emph{helical} gap can be opened by applying a magnetic field. However, previous work on the edge states of two-dimensional topological insulators has shown the possibility of a helical gap even without the application of a magnetic field, but instead due to two-particle backscattering.\cite{Orth+2015} Proximity coupling such a time-reversal invariant system to an $s$-wave superconductor then leads to the appearance of fourfold-degenerate zero-energy bound states, i.e., $\mathbb{Z}_4$ parafermions. \cite{Zhang+2014,Orth+2015,Peng+2016} This system hosts quasiparticles with fractionalized $e/2$ charge, which will result in an $8 \pi$ periodic component of the Josephson current in a weak-link S-N-S junction. Recently, we showed that an analogous \emph{spin-umklapp} scattering can open a helical gap without magnetic field in strongly interacting wires with RSOC.\cite{Pedder+2016} This prompts the question of whether an $8 \pi$-periodic Josephson current can be detected in Rashba wires as well.

Theoretical works have also turned to unproximitized wires with RSOC,\cite{Aseev+17, Schmidt+13,Meng+14,Klinovaja+14, Schmidt+16, Kornich+17} and experimental studies have investigated transport in these wires too. Without induced superconductivity, the system only has a partial gap, and so does not contain localized bound states. Recent data \cite{Schaepers+2017} shows a dip in the conductance of InAs wires with an applied magnetic field, consistent with the appearance of a helical gap. The same authors also find a similar signature without applied field, possibly arising from a partial spin-umklapp gap. We note that the theoretically-required interaction strengths have already been reached in similar quantum wires,\cite{Hevroni+2015} and that the opening of the spin-umklapp gap may also have been seen in other systems (e.g.~Ref.~[\onlinecite{Zumbuhl+2014}]).

In this paper, we investigate a measurement scheme for an $8 \pi$-periodic Josephson current in systems which support two-particle backscattering. This unusual periodicity of the Josephson current can be seen in Shapiro step experiments, even when we allow for a weak breaking of time reversal symmetry (TRS). Breaking TRS lifts the fourfold degeneracy, so we no longer have tunnelling of $e/2$ quasiparticles in the ground state. However, we will show that even allowing for dominant tunnelling of Cooper pairs and charge-$e$ quasiparticles, driven Shapiro step experiments can still be tuned to a regime where the disappearance of Shapiro steps reveals the presence of the $8\pi$-periodic term in the Josephson current.

The structure of this paper is as follows. In section 2, we introduce the specific model we choose to study - an interacting Rashba wire proximity coupled to a superconductor. We then bosonize this model in section 3, and present a renormalization group (RG) analysis for the flow of the system parameters. We find there exist regimes in which fourfold degenerate localized states can be found. In section 4, we describe the implications of the degenerate ground state for Josephson effect measurements, and suggest a definitive Shapiro step measurement to identify the $8\pi$ contribution to the Josephson current arising from these states, even when this contribution is sub-dominant. In section 5, we describe the situation in the absence of TRS breaking, and construct the $\mathbb{Z}_4$ exchange statistics of the localized states, described by non-Abelian, parafermionic operators. Finally, in section 6, we discuss the relevance of our work for experimental investigations.

\section{Model}
\begin{figure}[t]
	\centering
	\includegraphics[width=1\linewidth]{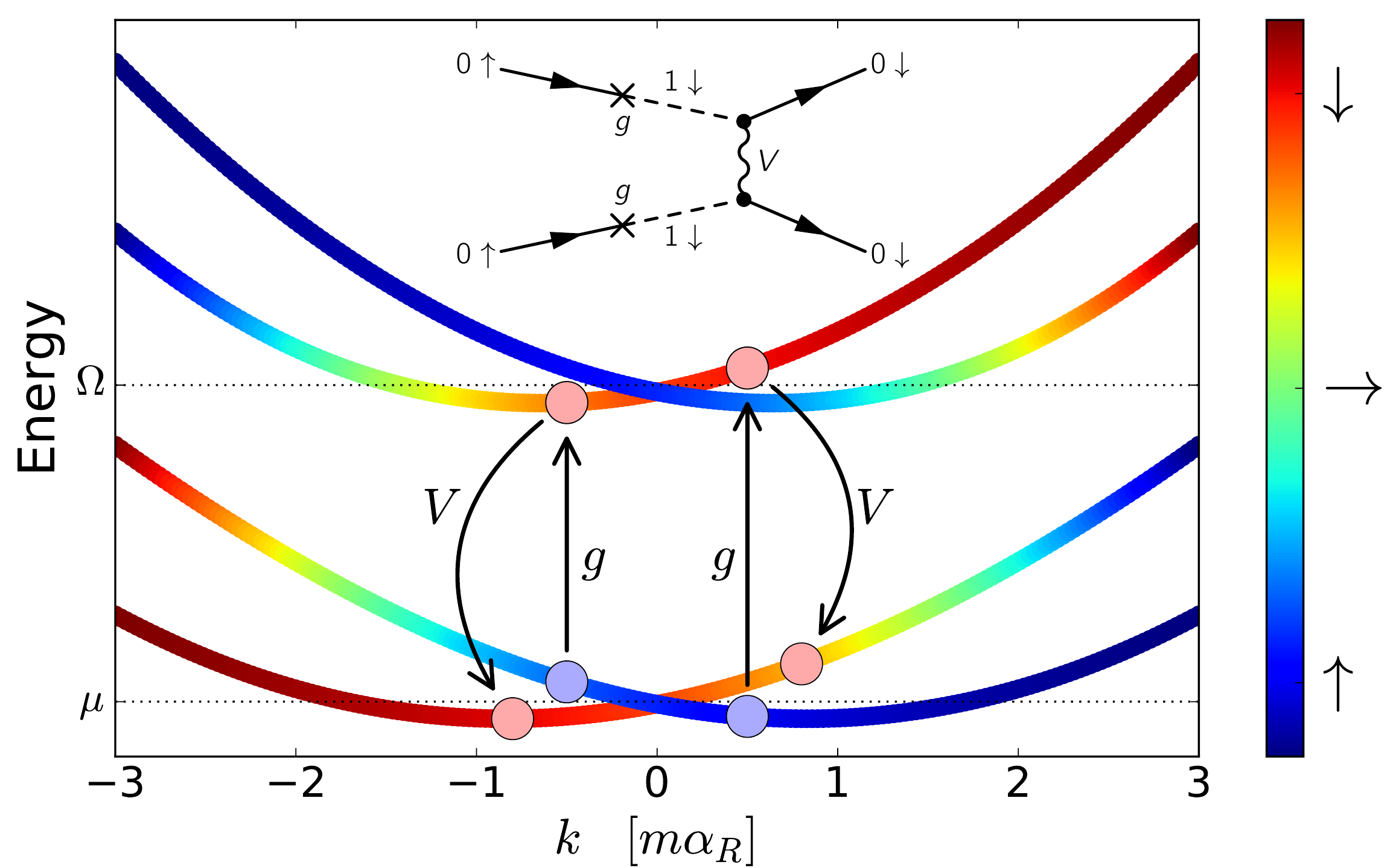}
	\caption{(color online) A plot of single-particle spectrum of a Rashba spin-orbit coupled nanowire showing the two lowest transverse sub-bands, whose spin character as a function of momentum is shown by the color-coding. Virtual transitions between sub-bands come with a spin-flip, and electron-electron interactions allow us to couple states in different sub-bands. The inset shows the exact process responsible for spin-umklapp scattering, where the virtual transition between sub-bands (denoted by $g$) changes the spin character of two incoming particles, and then the four-particle interaction vertex (denoted by $V$) couples these, resulting in a process in the lowest sub-band where two spin-$\uparrow$ particles are scattered into two spin-$\downarrow$ particles.}
	\label{spinumfig}
\end{figure}

We focus on the specific case of a long, quasi-1D nanowire aligned along the $x$ direction which is harmonically confined in the $y$ and $z$ directions. The interplay of the intrinsic spin-orbit coupling of a material and the breaking of inversion symmetry by a particular geometric arrangement, due to either the presence of a substrate, or to the application of an out of plane electric field, gives rise to RSOC.\cite{winkler_book} In the latter case, the strength of the electric field may be used to tune the magnitude of the RSOC, denoted as $\alpha_R$. The dynamics in the $z$-direction is not affected by the Rashba coupling and can be safely ignored. Thus, we can model a finite-width wire by using the following 2D Hamiltonian including RSOC with strength $\alpha_R$ \cite{moroz00,moroz00a,Starykh+2008}
\begin{equation}
	H = \frac{ p_x^2+p_y^2}{2m} + \frac{1}{2} m \omega^2 y^2 + \alpha_R (\sigma_x p_y - \sigma_y p_x),
	\label{RashbaHamiltonian}
\end{equation}
where $p_x$ and $p_y$ are the momentum components in the $x$- and $y$-directions, and $\sigma_{x,y}$ are Pauli matrices. The transversal confinement is modelled as a harmonic potential with frequency $\omega$. The system has translational invariance along the $x$-direction but is strongly confined along the $y$-direction, which leads to the appearance of higher excited bands separated from the lowest band by a spacing determined by the inverse width of the wire. As described in Ref.~[\onlinecite{Pedder+2016}], changes of sub-band come with a spin flip, which in tandem with a spin-conserving interaction can generate \emph{spin-umklapp} scattering when the chemical potential is at the Dirac point. This scattering converts e.g. two spin-up particles into two spin-down particles (see Fig.~\ref{spinumfig}).

\begin{figure*}[t]
	\centering
	\includegraphics[width=1\linewidth]{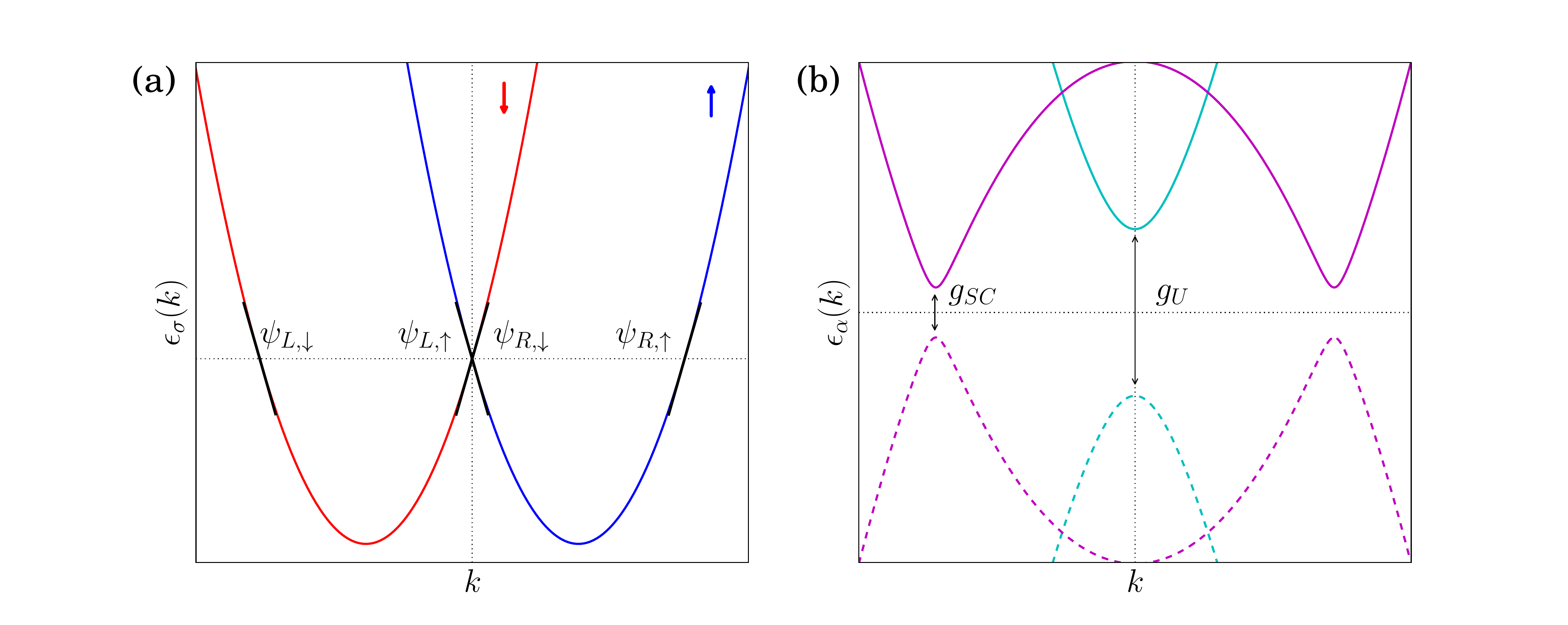}
	\caption{(color online) Panel (a) shows the position of the chemical potential and the four low-energy linearized modes used to bosonize the system. Panel (b) shows the reconstruction of the band structure as a result of the RG-relevant interaction processes between these modes. Umklapp scattering opens a helical gap of size $g_{\textrm{U}}$ between the two modes near $k=0$ and proximity-induced $s$-wave superconductivity opens gaps for the outer modes at $k=\pm 2m \alpha_R$ of size $g_{\textrm{SC}}$, resulting in a fully gapped system.}
	\label{Interactions}
\end{figure*}

Putting the chemical potential at the Dirac point leads to four low-energy modes at momenta $k=0,\pm k_F$ (where $k_F =2m \alpha_R$) as shown in Fig.~\ref{Interactions}. Correspondingly, for small energies, we can split the field operators up into four modes,
\begin{eqnarray}
	\psi_\uparrow (x) &\approx e^{ik_F x}  \psi_{R \uparrow}(x) + \psi_{L \uparrow}(x), \\
	\psi_\downarrow (x) &\approx \psi_{R \downarrow}(x) + e^{-ik_F x} \psi_{L \downarrow}(x).
\end{eqnarray}
Next, we express $\psi_{\alpha s}(x)$ for $\alpha \in \{L,R\}$ and $s \in \{ \uparrow,\downarrow \}$ in terms of its Fourier components,
\begin{equation}
	\psi_{\alpha s}(x) = \frac{1}{\sqrt{L}} \sum_k e^{i k x} \psi_{\alpha s,k},
\end{equation}
where $L$ is the length of the wire. In terms of these operators, the interaction Hamiltonian after projection to the lowest sub-band can be written as follows. The density-density interaction term reads\cite{Pedder+2016}
\begin{equation}
	V_\rho = \frac{\tilde{V}(0)}{L} {\sum_{\alpha,s}} \int \, dx \, \rho_{\alpha s} (x) \rho_{\alpha s} (x),
	\label{densityterms}
\end{equation}
where the fermionic densities are defined as usual as $\rho_{\alpha s}(x) = \psi^\dagger_{\alpha s}(x) \psi_{\alpha s}(x)$ and $\tilde{V}(q)$ is the Fourier transform of the interaction potential $V(\vec{r})$ projected to the lowest sub-band

Due to momentum conservation, the spin-flip terms only mix terms near $k=0$,
\begin{equation}
	V_{\rm{sf}} = v_{\rm{sf}} \int \, dx \left[ \psi^\dagger_{R\uparrow} (\partial_x \psi^\dagger_{R\uparrow}) (\partial_x \psi_{L\downarrow}) \psi_{L\downarrow} + \rm{h.c.} \right],
	\label{SpinFlip}
\end{equation}
where $v_{\rm{sf}}$ is a dimensionful parameter characterizing the strength of the spin-umklapp scattering.

Most of the spin-exchange terms arising from the Coulomb interaction can be expressed as density-density interactions, leading merely to changes in the coefficients of the terms in Eq.~(\ref{densityterms}). We separate out the single non-density-density term
\begin{equation}
	V_{\rm{S}} = v_{\rm{S}} \int \, dx \biggl( \psi_{L\downarrow}^\dagger \psi_{R\uparrow}^\dagger \psi_{L\uparrow} \psi_{R\downarrow} + \rm{h.c.}  \biggr),
	\label{SDWTerm}
\end{equation}
which corresponds to an interaction between inner and outer bands with strength $v_{\rm{S}}$. In the limit in which this term dominates, it results in a spin-density wave state at $q=2k_F$, and has been discussed in detail in Ref.~[\onlinecite{Starykh+2008}].

Finally, we allow for the possibility of proximity-induced coupling to an $s$-wave superconductor, which pairs spin-up and spin-down electrons, and so has an effect at all the Fermi points in Fig.~\ref{Interactions}a. This pairing contribution to the Hamiltonian is
\begin{equation}
	V_{\rm{SC}}= v_{\rm{SC}} \int \, dx \left( \psi_{R\uparrow}^\dagger \psi_{L\downarrow}^\dagger + \psi_{L\uparrow}^\dagger \psi_{R\downarrow}^\dagger + \rm{h.c.} \right),
\end{equation}
where $v_{\rm{SC}}$ is determined by the strength of the proximity-coupling to the superconductor. We now analyse the competition between these three possible interaction channels, parametrised by $v_{\rm{sf}}$, $v_{\rm{S}}$ and $v_{\rm{SC}}$.

\section{Bosonization \& renormalization group analysis.}
To further analyze the interacting system, we write the Hamiltonian in terms of bosonic operators $\phi_{\pm}$ and $\theta_{\pm}$ by defining
\begin{align}
	\psi_{R \uparrow} &=  \frac{\eta_{R \uparrow} }{\sqrt{2\pi a}} e^{-i(\phi_+ - \theta_+)}, \hspace{5mm} \psi_{R \downarrow} =  \frac{ \eta_{R \downarrow}}{\sqrt{2\pi a}} e^{-i(\phi_- - \theta_-)},\nonumber \\
	\psi_{L \uparrow} &= \frac{\eta_{L \uparrow} }{\sqrt{2\pi a}}  e^{i(\phi_- + \theta_-)}, \hspace{5mm} \psi_{L \downarrow} =  \frac{ \eta_{L \downarrow} }{ \sqrt{2\pi a}} e^{i(\phi_+ + \theta_+)},
\end{align}
where $\eta_{\alpha \sigma}$ are Klein factors and $a$ denotes the short-distance cutoff, related to the large momentum cutoff $\Lambda$ by $a \sim \Lambda^{-1}$. Here, $\phi_+(x)$ and $\theta_+(x)$ are canonically conjugate bosonic operators for degrees of freedom near $k = \pm k_F$ where $k_F=2m\alpha_R$, whereas $\phi_-(x)$ and $\theta_-(x)$ describe modes near $k=0$. In these variables, the Hamiltonian consists of two Luttinger Hamiltonians for the $+$ and $-$ species, with approximately equal Luttinger parameters $K_\pm$, and interaction terms reflecting Eqs.~(\ref{SpinFlip}) and (\ref{SDWTerm}). Moreover, one obtains derivative terms which couple the two species $\kappa_\phi \partial_x \phi_+ \partial_x \phi_-$ and $\kappa_\theta \partial_x \theta_+ \partial_x \theta_-$.

Following Ref.~[\onlinecite{Starykh+2008}], we diagonalize the quadratic parts of the Hamiltonian by going to the charge-spin basis $\phi_{\rho,\sigma} = (\phi_+ \pm \phi_-)/\sqrt{2}$ and $\theta_{\rho,\sigma} = (\theta_+ \pm \theta_- )/\sqrt{2}$ so that
\begin{equation}
	H_0 = \sum_{a=\rho,\sigma} \frac{v_a}{2 \pi} \int \, dx \left[ \frac{(\partial_x \phi_a)^2}{K_a} + K_a (\partial_x \theta_a)^2 \right],
	\label{NonInteractingHamiltonian}
\end{equation}
where $v_{\rho,\sigma}$ are the respective sound velocities of the modes. For repulsive interactions, we have $K_\rho < 1$ and $K_\sigma < 1$ \cite{Starykh+2008}. In addition to $H_0$, we obtain the two competing interaction terms,
\begin{align}
	V_{\rm{S}} &= \frac{g_{\rm{S}}}{(2 \pi a)^2} \int \, dx \, \cos [2\sqrt{2} \theta_\sigma], \\
	V_{\rm{sf}} &= \frac{ g_{\rm{U}}}{(2 \pi a)^2} \int \, dx \, \cos [ 2\sqrt{2}(\phi_\rho-\phi_\sigma)].
\end{align}
Proximity-induced $s$-wave superconductivity gives a contribution which reads in bosonized form,
\begin{equation}
	V_{\rm{SC}} = \frac{ g_{\rm{SC}}}{(2 \pi a)^2} \int \,dx \, \biggl( \cos[\sqrt{2}(\theta_\rho + \theta_\sigma)] + \{\theta_\sigma \rightarrow -\theta_\sigma\} \biggr).
\end{equation}
For weak interactions, the parameters of the model can be determined precisely in the bosonization procedure (see Ref.~[\onlinecite{Pedder+2016}]). However, for strong interactions it is more convenient to regard the parameters $v_{\rho,\sigma}$ and $K_{\rho,\sigma}$ as well as the three coupling strengths $g_{\rm{S}}$, $g_{\rm{U}}$ and $g_{\rm{SC}}$ as effective parameters, which may flow independently under renormalization as we change the cutoff $a$.

We calculate the flow of the various coupling constants using real-space RG calculation based on operator product expansions.\cite{cardy96} We find the following first-order RG equations for the coupling constants of the cosine terms
\begin{align}
	\frac{d g_{\rm{S}}}{d \ell} &= \left( 2-\frac{2}{K_\sigma} \right) g_{\rm{S}}, \label{RGsdw}  \\
	\frac{d g_{\rm{U}}}{d \ell} &=  2(1-K_\sigma-K_\rho) g_{\rm{U}}, \label{RGumklapp}     \\
	\frac{d g_{\rm{SC}}}{d\ell} &= \left( 2-\frac{1}{2K_\sigma} - \frac{1}{2K_\rho} \right) g_{\rm{SC}}, \label{RGsuperconductivity}
\end{align}
implying that the spin-density wave term is always irrelevant for repulsive interactions ($K_\sigma < 1$).\cite{Starykh+2008} The spin-umklapp term, by contrast, can become relevant for strong interactions where $K_\rho + K_\sigma < 1$. Finally, the superconducting term is relevant for $K_\rho^{-1} + K_\sigma^{-1} < 4$.

We would like to point out that for $\alpha_R =0$, the system becomes $\rm{SU}(2)$ invariant. In that case, the spin-umklapp term vanishes because spin is conserved and one finds the well-known Kosterlitz-Thouless RG flow which brings $K_\sigma \to 1$ as $g_{\rm{S}} \to 0$. In contrast, for $\alpha_R \neq 0$, $K_\sigma$ is not constrained and strong repulsive interactions lead to $K_\sigma \ll 1$.

We start the RG flow from an initial value $a = a_0$ and flow towards $a \sim L$, the length of the wire. Generically, the RG flow will stop at a finite value $a_\infty < L$ as soon as one of the dimensionless coupling constants $g_{\rm{SC,U}}$ approaches one. The bare value of $g_{\rm{SC}}(a_0)$ is determined by the strength of the proximity coupling to the superconductor, which can be experimentally optimized. The bare value $g_{\rm{U}}(a_0)$ depends on the separation between the lowest sub-bands, and so depends on the transverse confinement (i.e.~the physical width) of the wire.

To generate zero-energy bound states, spin-umklapp scattering must gap out the modes near $k=0$, whereas proximity-induced superconductivity should open a gap for the modes at $k=\pm k_F$, see Fig.~\ref{Interactions}(b). Superconductivity affects all modes, so this is only possible if at the end of the RG flow $|g_{\rm{U}}(a_\infty)| > |g_{\rm{SC}}(a_\infty)| > 0$. Strong electron-electron interactions result in $K_\rho <1/2$ and $K_\sigma<1/2$, which a priori makes the spin-umklapp term relevant and the superconducting term irrelevant. However, since the RG flow is cut off at a finite length scale, one will generally find a non-zero $|g_{\rm{SC}}(a_\infty)| > 0$ at the end of the RG flow, meaning that a superconducting gap will still open.

\section{Josephson effect \& Shapiro steps.}

A zero-bias conductance peak is a possible experimental signature of localized Majorana fermions associated to a helical gap. However, such peaks could arise from other mechanisms, e.g. disorder,\cite{Pikulin+2012,Liu+2012} and do not directly indicate the degeneracy of the states involved. In particular, the umklapp gap, and associated fractionally-charged states we previously proposed in Ref.~[\onlinecite{Pedder+2016}] would lead to a similar zero-bias anomaly in a system with proximity-induced superconductivity as that caused by Majorana fermions, albeit at vanishing magnetic field. To uniquely discriminate these particular bound states, we instead propose to discover their presence via the periodicity of the Josephson effect, similar to the corresponding proposal for Majorana fermions.\cite{Fu+2009,rokhinson12}

\begin{figure}[t]
	\centering
	\includegraphics[width=0.95\linewidth]{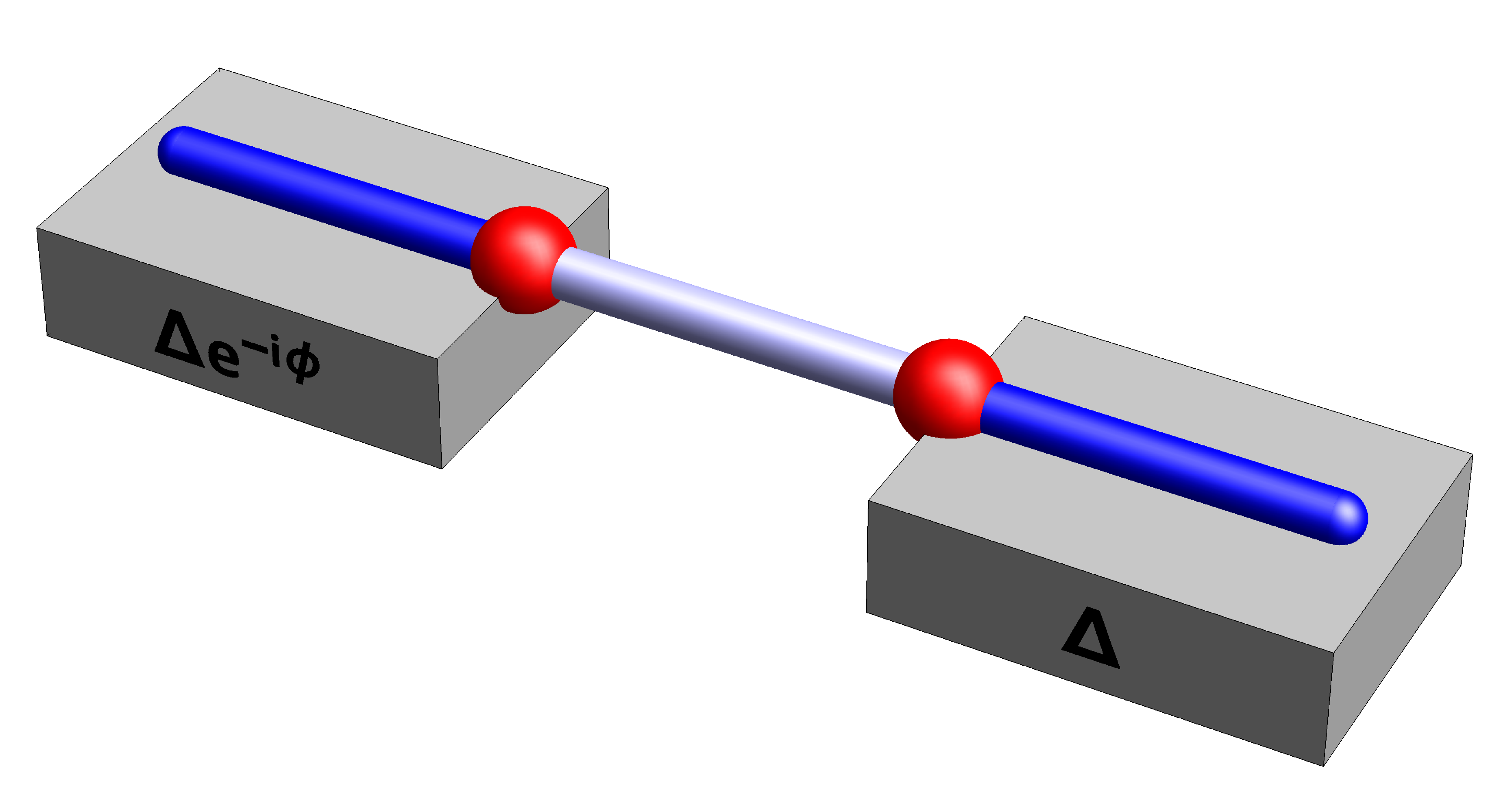}
	\caption{Experimental setup for the measurement of the $8 \pi$ periodic Josephson effect. Two superconductors underneath a Rashba wire are held at a phase difference $\phi$. The blue portions of the wire are gapped by superconductivity, whereas the grey section is gapped by umklapp scattering. Fractionally charged bound states are indicated by red spheres.}
	\label{SCSetup}
\end{figure}

To investigate the role of fractionally-charged tunnelling states on the Josephson effect, we follow the logic of Refs.~[\onlinecite{Zhang+2014,Orth+2015}] and consider an arrangement with two superconducting contacts with phase difference $\phi$ placed under a Rashba wire partially gapped by spin-umklapp scattering (see Fig.~\ref{SCSetup}), in an analogous arrangement to the experimental setup of Ref.~[\onlinecite{rokhinson12}]. In the absence of TRS breaking, the wire adjacent to the edges of the superconductors will host zero-energy modes with charge $e/2$, which will dominate the transport at low energies and for a short junction.\cite{Pedder+2016} Tunnelling of a single quasiparticle through the junction will change the parity of the end states. In order to satisfy the boundary conditions due to the applied superconducting phase difference, \emph{four} $e/2$ quasiparticles must tunnel via the bound states, leading to an $8\pi$ periodic Josephson effect \cite{Zhang+2014}. The TRS breaking caused by the superconducting phase difference causes a slight lifting of the fourfold degeneracy, but for realistic parameters this shift is negligible.\cite{Zhang+2014}

\begin{figure*}[t]
	\centering
	\includegraphics[width=\linewidth]{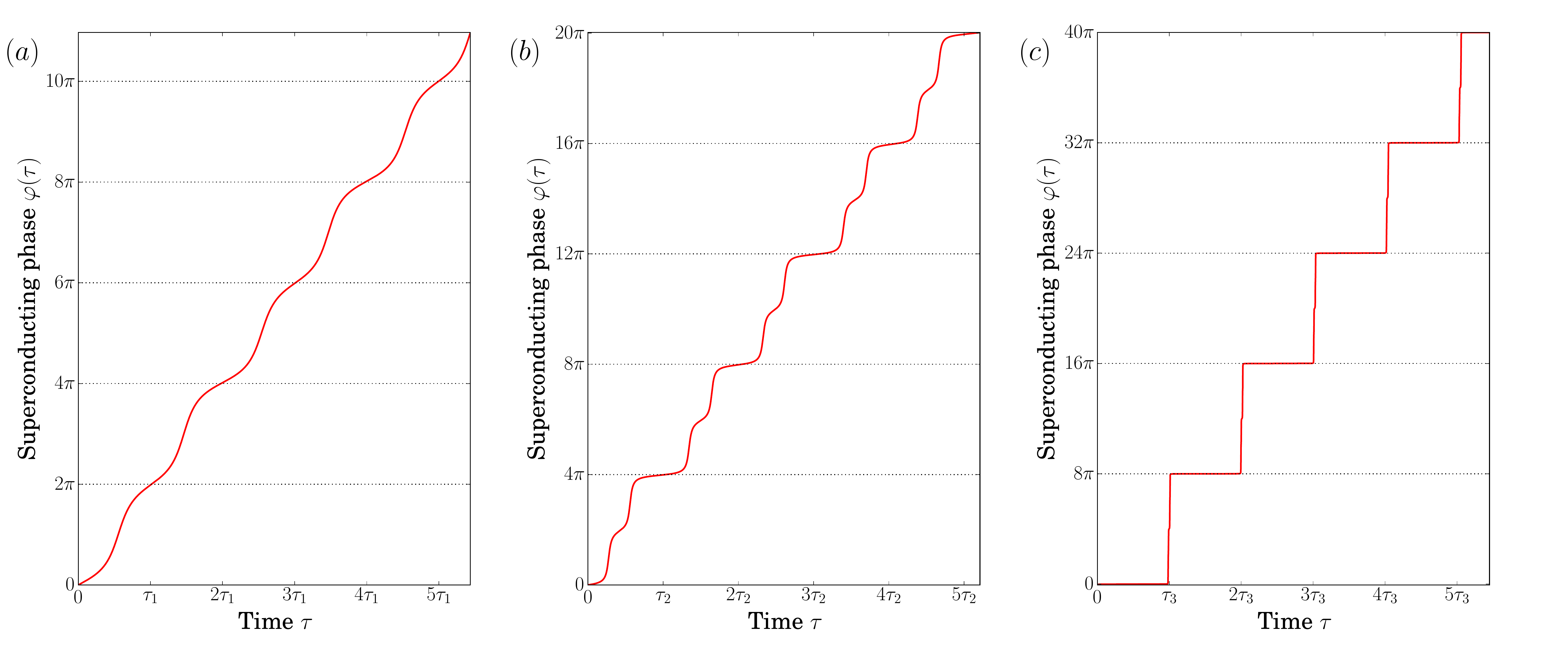}
	\caption{Numerical solutions to Eq.~(\ref{Jcurrent2}) with a d.c. bias current ($\alpha_1=0$), showing different periodicity behaviour depending on the amplitude of the d.c. current. We have made the generic choice of $\alpha_m=1/10$ and $\alpha_p=1/100$ for the relative size of the contributions to the Josephson current from Majorana fermions and parafermions. Panel (a) shows the superconducting phase changing in $2\pi$ steps for the choice $\alpha_0=1.5$, whereas panel (b) is plotted for $\alpha_0=1.1$, and shows residual small steps at $2\pi$ and dominant steps at $4\pi$ which corresponds to the dominant transport being via tunneling of Majorana fermions. In panel (c), we choose $\alpha_0=1.08$, and recover jumps of $8\pi$ in the superconducting phase due to tunnelling of parafermions.}
	\label{FigNoDriveCurrent}
\end{figure*}

In real material samples, TRS may also be weakly broken by magnetic impurities, thereby lifting the fourfold degeneracy to a twofold one. In our finite-length Rashba wire, there will generically be a non-zero overlap of the modes at each end of the wire and so the degeneracy between the modes will be split, although this effect is exponentially suppressed in the length of the wire. In any case, the un-driven Josephson current is no longer $8\pi$ periodic. This raises the question of whether remnants of the $8 \pi$ periodicity can be observed in such a non-ideal setting.

A possible answer was proposed for similar problems in Majorana nanowires,\cite{Dominguez+2012} where a $4 \pi$ periodicity is reduced by parity-flipping perturbations to a trivial $2\pi$ periodicity: by driving the current in the junction at a finite frequency, we allow Landau-Zener tunnelling between the different low-lying states. Then Shapiro step measurements \cite{Shapiro+1963} can still distinguish higher periodic components even when those are very weak, as in the case of a $4\pi$ periodicity recently reported in experiments on Majorana bound states \cite{rokhinson12, Wiedenmann+2015}.

In the case of a driven junction, Landau-Zener tunnelling gives us access to all the low-energy modes, even when they are subject to a small splitting. Allowing for the possibility of Josephson tunnelling of Cooper pairs, as well as for the tunnelling of charge $e$ and charge $e/2$ quasiparticles through the weak link in our system, the total Josephson current flowing is given by
\begin{equation}
	I(\varphi(t)) = i_c \sin [\varphi(t)] +i_m \sin \left[\frac{\varphi(t)}{2}\right] + i_p \sin \left[\frac{\varphi(t)}{4}\right].
\end{equation}
The amplitude $i_c$ accounts for the current due to the tunnelling of Cooper pairs (this is the critical current above which the junction becomes ``normal''). The parameters $i_m$ and $i_p$ similarly account for the tunnelling of fractionalized charge $e$ and charge $e/2$ quasiparticles through the junction. Note that these contributions to the Josephson current are periodic under shifts $\varphi(t) \rightarrow \varphi(t)+2 \pi$, $\varphi(t) \rightarrow \varphi(t)+4 \pi$ and $\varphi(t) \rightarrow \varphi(t)+8 \pi$ respectively. The Josephson equation relating the rate of change of the superconducting phase $\varphi$ to the voltage across the junction reads
\begin{equation}
	\dot{\varphi}(t) = \frac{2e}{\hbar} V(t).
\end{equation}
We \emph{current-bias} our Josephson junction using a constant current with a small a.c.~component $I_0 + I_1 \sin (\omega t)$. The current through the junction consists of two parallel components, a \emph{tunnel current} given by $I(\varphi(t))$ and a \emph{resistive current} due to ohmic quasiparticle transport in the junction. Equating the sum of these two contributions to the bias current, the gauge-invariant phase $\varphi(t)$ is described by the dynamical equation
\begin{equation}
	I_0 + I_1 \sin (\omega t) = i_c \sin [\varphi(t)] +i_m \sin \left[\frac{\varphi(t)}{2}\right] + i_p \sin \left[\frac{\varphi(t)}{4}\right] + \frac{\hbar  \dot{\varphi}(t)}{2eR}.
	\label{Jcurrent}
\end{equation}
 Writing Eq.~(\ref{Jcurrent}) in terms of the rescaled variables $\tau=2eR i_c t/\hbar$, and $\tilde{\omega} = \hbar \omega/2eR i_c$, we find the equation
\begin{equation}
	\dot{\varphi}(\tau) = \alpha_0 + \alpha_1 \sin (\tilde{\omega} \tau) -\sin [\varphi(\tau)] -\alpha_m \sin \left[\frac{\varphi(\tau)}{2}\right]- \alpha_p \sin \left[\frac{\varphi(\tau)}{4}\right]
	\label{Jcurrent2}
\end{equation}
This equation must be solved numerically.

In order to see that the $8\pi$ periodicity can win out, even when $\alpha_c=1>\alpha_m>\alpha_p$, we first solve Eq.~(\ref{Jcurrent2}) without the a.c. drive current, i.e. $\alpha_1=0$ (see Fig.~\ref{FigNoDriveCurrent}). For a generic choice of the d.c. bias current, $\alpha_0$, the superconducting phase is $2\pi$ periodic, reflecting the dominance of the Cooper pair tunnelling (Fig.~\ref{FigNoDriveCurrent}a). However, we find that as we approach a critical value of $\alpha_0$, we first see the $4\pi$ periodic term due to the charge-$e$ quasiparticles (Fig.~\ref{FigNoDriveCurrent}b) and then the $8\pi$ contribution from the charge-$e/2$ quasiparticles (Fig.~\ref{FigNoDriveCurrent}c) become dominant.

\begin{figure*}[t]
	\centering
	\includegraphics[width=0.95\linewidth]{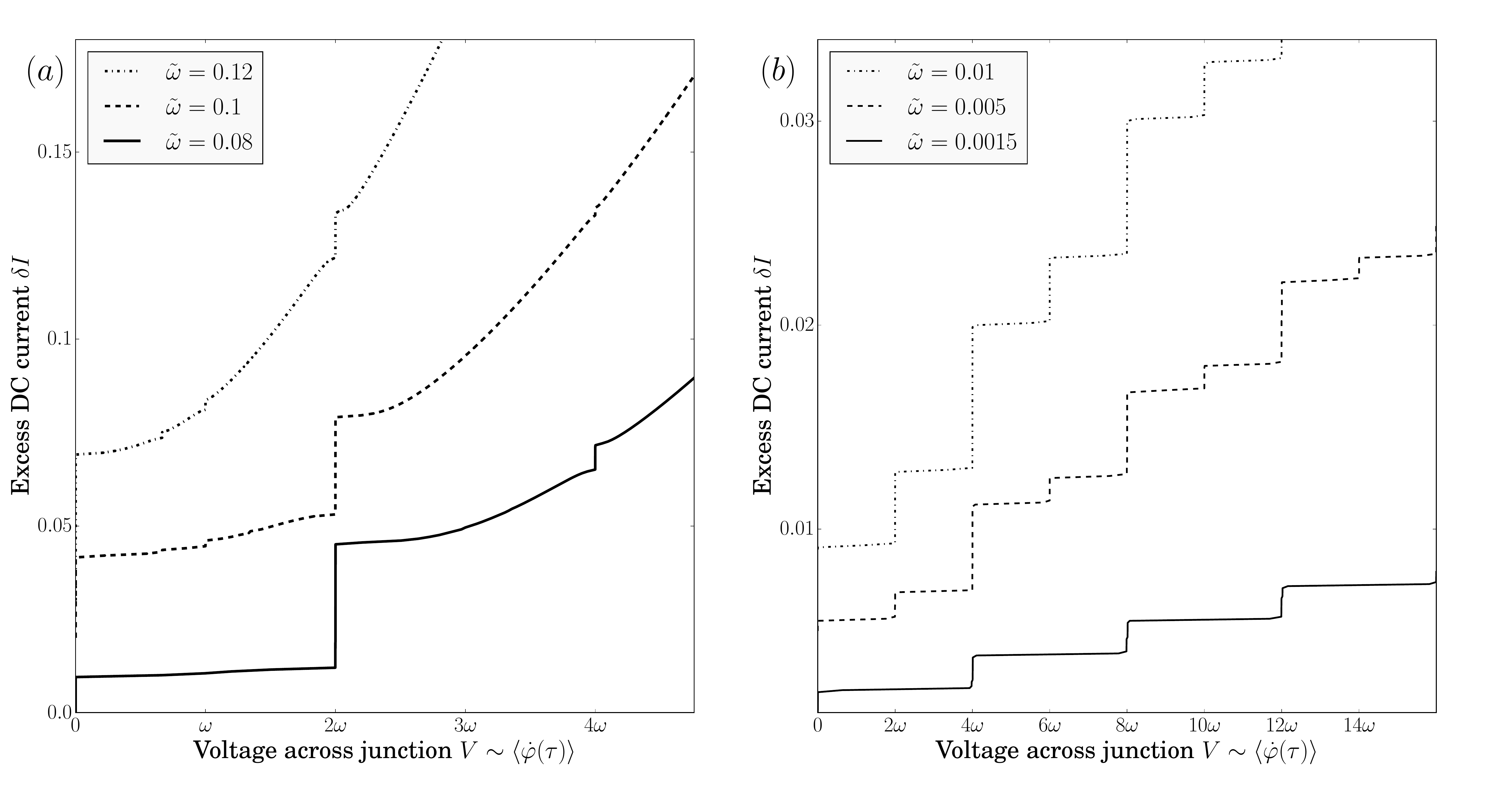}
	\caption{A plot of the current-voltage relationship for the system driven with a small amplitude a.c.~current with a large d.c.~offset (with corresponding magnitudes $\alpha_0=1.08+\delta I$ and $\alpha_1=0.01$). The excess current $\delta I$ over the critical value at zero drive is plotted on the vertical axis. The Josephson junction has a dominant $2\pi$ periodic component, and sub-leading $4\pi$ and $8\pi$ periodic components ($\alpha_m=1/10$ and $\alpha_p=1/100$). (a) At the fastest drive frequency (comparable to that of the $2\pi$ periodic component of the Josephson current) shown in the top curve, we find all integer Shapiro steps. As we drive at a slower frequency (middle curve), the first Shapiro step shrinks, and then disappears, leaving only Shapiro steps which are multiples of 2. (b) Driving at smaller frequencies, we find a double-step structure (top curve), but the size of the intermediate steps shrink as we reduce $\tilde{\omega}$ (middle curve). Eventually, at the lowest frequencies, we see a $4n \omega$ step structure. Note that the extra steps at $V \approx 2\omega/3$ in the top two lines in panel (a) result from the fact that we are driving non-adiabtically, but are not relevant to experimental measurements as they do not occur at integer multiples of the driving frequency.}
	\label{shapirostepsfig}
\end{figure*}

The winding up of the superconducting phase is not a property which is easy to directly measure, so to give an experimentally accessible measurement, we must drive the system to see the degeneracy of the low-lying modes. To do this, we switch on the small a.c. component to the bias current $\alpha_1$ (experimentally achieved by irradiating the junction with microwaves). Tuning the driving frequency allows us to access three distinct regimes, in which steps occur in the experimentally-measurable I-V curves for the junction at distinct multiples of the driving frequency $\tilde{\omega}$. For high frequencies, we find Shapiro steps are present at \emph{all} integer multiples of the driving frequency, indicating that the transport is dominated by conventional Josephson tunnelling of Cooper pairs through the junction. Reducing the frequency of the drive, we recover the results of Ref.~[\onlinecite{Dominguez+2012}], that only the \emph{odd}-numbered Shapiro steps remain. Finally, for lower frequencies still, there exists a regime in which only every $(4n)^{th}$ Shapiro step survives (see Fig. \ref{shapirostepsfig}).

The appearance of \emph{extra} steps can be caused e.g.~by disorder, or strongly non-adiabatic driving. However, mechanisms other than the one described by which Shapiro steps may \emph{disappear} seem to be unknown. As the disappearing Shapiro steps are robust even when the $2\pi$ and $4\pi$ contributions to the Josephson current are dominant over the $8 \pi$ component, they therefore provide a highly selective test of the existence of an interaction-generated helical gap, even in the presence of weak TR symmetry breaking.

\section{Bound states}

Several works \cite{Fidkowski+2011,turner11} have suggested that in a spinless, one-dimensional system, the greatest achievable topological degeneracy is twofold, leading to the statement that only Majorana fermions can exist in one-dimensional systems. Our system does not contradict this theorem because in our case the ground state degeneracy is not entirely topological. Indeed, it can be viewed as a twofold topological degeneracy combined with a two-fold degeneracy due to TRS \cite{Sela+2011}. This second degeneracy can be lifted by local TRS-breaking perturbations such as a magnetic field. In that case, only the topological part of the ground state degeneracy survives, and one recovers the $4\pi$ periodicity of the Josephson effect seen for Majorana bound states.\cite{Fu+2009}

The Shapiro step measurement suggested in the previous section will demonstrate the existence of a spin-umklapp generated gap even in disordered nanowires which may have weak TRS breaking. However, the level of disorder in current state-of-the-art nanowires and nanowire junctions is so low as to demonstrate ballistic electron transport.\cite{Schaepers+2017,Kouwenhoven+2017a,Kouwenhoven+2016} In such clean wires, and without explicit TRS breaking, it may even be possible to see the existence of degenerate, localized zero-energy bound states. To investigate the form of these bound states, we use the unfolding transformation described in Refs.~[\onlinecite{Oreg+2014,Giamarchi+2004}]. This transformation can be used to map our system with length $L$ and open boundary conditions to a system of length $2L$ and periodic boundary conditions. Explicitly, we construct the unfolded \emph {chiral} fields
\begin{align}
	\xi_R(\tx) = \left\{\begin{array}{ll}
		\varphi_{R+} (\tx) & {\rm{for}} \  0 \leq \tx \leq L\\
		\varphi_{L-} (2L-\tx) & {\rm{for}} \,\, L \leq \tx \leq 2L
	\end{array}\right.\\
	\xi_L (\tx) = \left\{\begin{array}{ll}
		\varphi_{L+} (\tx) & {\rm{for}} \ 0 \leq \tx \leq L \\
		\varphi_{R-} (2L-\tx) & {\rm{for}} \ L \leq \tx \leq 2L
	\end{array}\right.
\end{align}
where $\varphi_{\alpha \nu} = \alpha \phi_{\nu} - \theta_{\nu}$ with $\alpha=R,L$ and $\nu = \pm$. In order that the fermionic fields obey the vanishing boundary conditions at $x=0$ and $x=L$, we find that the bosonic fields must obey
\begin{align}
	\varphi_{L+} (0) = \varphi_{R-}(0), & \qquad \varphi_{L+} (L) = \varphi_{R-}(L)\nonumber \\
	\varphi_{L-} (0) = \varphi_{R+}(0), & \qquad \varphi_{L-} (L) = \varphi_{R+}(L)
\end{align}
Note that in this transformation, the degrees of freedom $(\phi_+,\theta_+)$ are mapped on the range $\tx \in [0,L]$, whereas the $(\phi_-,\theta_-)$ fields are mapped on the range $\tx \in [L,2L]$. Since the original chiral fields satisfy $[\varphi_{\alpha\nu}(x),\varphi_{\alpha^\prime\nu^\prime}(x')] =  i \pi \alpha\delta_{\alpha \alpha^\prime} \delta_{\nu \nu^\prime} {\rm{sgn}}(x-x')$, we find that the unfolded fields obey the correct chiral commutation relations
\begin{equation}
	[\xi_{\alpha}(\tx),\xi_{\alpha'}(\tx')] = i\pi \alpha \delta_{\alpha\alpha'}{\rm{sgn}}(\tx-\tx'),
\end{equation}
on the whole interval $\tx,\tx' \in [0,2L]$. In terms of these unfolded fields, our Hamiltonian for the relevant perturbations arising from umklapp scattering and superconductivity reads
\begin{align}
	V_{\rm{sf}} + V_{\rm{SC}} &=& \int_0^{2L} d\tx \biggl[ g_{\rm{U}}(\tx) \cos(2[\xi_R(\tx)-\xi_L(\tx)]) \nonumber \\
	& & \hspace{12mm} + g_{\rm{SC}}(\tx) \cos(\xi_R(\tx)+\xi_L(\tx)) \biggr]
\end{align}
where the position-dependent couplings $g_{\rm{U}}(\tx)$ and $g_{\rm{SC}}(\tx)$ have support on $\tx\in [0,L]$ and $\tx\in [L,2L]$ respectively. The unfolded system consists of two adjacent regions. Between $\tx=0$ and $\tx=L$, we have a region of superconductor where the field $\theta(\tx)=-(\xi_R+\xi_L)/2$ is pinned by the term $\cos[2\theta]$ to the value $\theta^\star = (n+1/2)\pi$ for integer $n$. From $\tx=L$ to $\tx=2L$, there is a region of ``Mott insulator'' where the field $\phi=(\xi_R-\xi_L)/2$ is pinned by the term $\cos[4 \phi]$ to the value $\phi^\star=1/2(m+1/2)\pi$ for integer $m$. The spectrum is completely gapped, except possibly at the boundaries between the two regions, $\tx = 0$ and $\tx = L$, where the parafermion states we describe emerge. The unfolded system is identical to the topological insulator edge state system studied in Ref.~[\onlinecite{Orth+2015}], which tells us that our original system contains $\mathbb{Z}_4$ parafermion state at its ends.

Let us reproduce the essential parts of the derivation here. We define the total charge and total spin operators for the system, according to
\begin{align}
	\pi S &= \theta(2L)-\theta(0), \\
	\pi Q &= \phi(2L)-\phi(0).
\end{align}
Despite the fact that the fermionic fields must be continuous, the bosonic fields $\phi$ and $\theta$ may jump by integer multiples of $2\pi$, so that $S$ and $Q$ can be non-zero in spite of the periodic boundary conditions. The spin $S$ of the system takes integer values, and is conserved $mod(4)$ so  $s=\{0,1,2,3\}$ (measured in units of $\hbar/2$), whereas the charge $Q$ takes half-integer values and is conserved $mod(2)$ so $q=\{0,\frac{1}{2},1,\frac{3}{2} \}$ (measured in units of $e$). Since we may only add integer amounts of electronic charge to our junction, we must restrict our value of the charge to be $q\in \{0,1\}$. Note that in a system where there are several junctions between superconducting and Mott insulating regions, it is perfectly acceptable to have states with half-integer charge, as long as the total charge of the complete system is restricted again to $q\in \{0,1\}$. The state of our system is then defined by $|s,q \rangle$. Since every physical electron carries one unit of spin, this means that for the charge state $q=0$, only the two total spin states $s\in \{0,2\}$ are permissible. Similarly, $q = 1$ requires $s \in \{1,3\}$. Hence, we have a total fourfold degeneracy of the ground state.

To see explicitly the parafermionic statistics of the bound states at $x=0$ and $x=L$, we write the pinned values of the fields in terms of integer-spectrum operators $m,n_1,n_2$ as
\begin{eqnarray}
	\phi_1 &=& \frac{1}{2} \left( \pi m + \frac{\pi}{2} \right), \nonumber \\
	\theta_{1,2} &=&  \pi n_{1,2} + \frac{\pi}{2}.
\end{eqnarray}
These operators then have commutation relations
\begin{equation}
	[m,n_1] = -\frac{i}{\pi},
\end{equation}
\begin{equation}
	[m,n_2] = 0.
\end{equation}
The total spin operator is given by
\begin{equation}
	\hat{S} = e^{i \pi m/2} = e^{-i\pi/4} e^{i \phi_1}
\end{equation}
and the total spin in the system is $\langle \hat{S} \rangle = (\theta_1-\theta_2)/\pi = n_2-n_1$. Then the parafermion states obeying
\begin{equation}
	\chi_1 \chi_2 = e^{-i\pi/2} \chi_2 \chi_1
\end{equation}
are given by
\begin{align}
	\chi_1 &= T_Q e^{i \pi m/2}, \notag \\
	\chi_2 &= e^{i\pi/4} T_Q e^{i \pi m/2} e^{i \pi (n_2-n_1)/2},
\end{align}
where $T_Q$ is the raising operator for charge $T_Q |s,q \rangle = |s,q+1 \rangle$.

\section{Discussion.}

The opening of the spin-umklapp gap is rather generic. In a wire with RSOC and electron-electron interactions this gap will lead to a reduction of the normal-state conductance of from $2e^2/h$ to $e^2/h$ as the chemical potential approaches the Dirac point, which may already have been seen in InAs nanowires.\cite{Schaepers+2017} We would like to point out that the required strong interactions have already been seen in these wires.\cite{Hevroni+2015} Umklapp scattering has also recently been invoked to explain the reduction in conductance observed in InAs/GaSb topological insulator edge states.\cite{li15} Introducing a weak superconducting proximity effect in either these wires or edge states will then lead to the creation of bound states and allow the observation of the $8\pi$ periodic Josephson current component.

Our system is susceptible to bulk disorder. Firstly, disorder causes one-particle backscattering which corresponds to another cosine term in the bosonized Hamiltonian. For umklapp scattering to win out, we require that the amplitude of the umklapp term is larger than the disorder potential at the end of the RG flow. Moreover, disorder can cause local deviations of the chemical potential away from $\mu = 0$. If these fluctuations are small compared to the umklapp gap, the bound states will persist. Recent measurements on InAs wires \cite{Schaepers+2017} and InSb wires and arrays \cite{Kouwenhoven+2017a,Kouwenhoven+2016} show ballistic transport which is indicative of very low levels of bulk disorder. Weak magnetic disorder will result in explicit breaking of time-reversal symmetry, and so the localized $\mathbb{Z}_4$ parafermionic bound states associated to the charge $e/2$ quasiparticles will cease to exist, but the sub-leading $8\pi$-periodic term will still be visible in Shapiro step experiments.

To summarize, we propose an experimental scheme allowing us to observe an $8\pi$ periodic Josephson current occurring via the tunnelling of fractional $e/2$ charges through the junction. The associated Shapiro step structure in which only the $4n$ steps survive (where $n \in \mathbb{Z}$) would be a definitive signature of fourfold degenerate bound states, associated to an umklapp gap generated by two-particle backscattering in strongly interacting Rashba wires or edge states of two-dimensional topological insulators. In the absence of TRS breaking perturbations, the bound state degeneracy is protected by a combination of time-reversal symmetry and fermion parity symmetry. The operators describing these bound states in the ground state manifold satisfy a $\mathbb{Z}_4$ parafermionic algebra.

\emph{Acknowledgements.} TLS and CJP are supported by the National Research Fund, Luxembourg under grant ATTRACT 7556175. TM is funded by Deutsche Forschungsgemeinschaft through GRK 1621, SFB 1143, and the Emmy-Noether program ME 4844/1. RPT acknowledges financial support from the Swiss National Science Foundation.

\end{document}